# Unveiling the Direct Piezoelectric Effect on Piezo-phototronic Coupling in Ferroelectrics: First Principle Study Assisted Experimental Approach

Koyal Suman Samantaray[1], Sourabh Kumar[1], P Maneesha [1], Dilip Sasmal[1], Suresh Chandra Baral[1], B.R. Vaishnavi Krupa[2], Arup Dasgupta[2], K Harrabi[3,4], A Mekki[3,5], and Somaditya Sen[1*]

[1]*Department of Physics, Indian Institute of Technology Indore, Indore, 453552, India*

[2]*Physical Metallurgy Division, Indira Gandhi Centre for Atomic Research, HBNI, Kalpakkam, 603102, Tamil Nadu, India*

[3]*Department of Physics, King Fahd University of Petroleum and Minerals, Dhahran 31261, Saudi Arabia*

[4]*Interdisciplinary Research Center (RC) for Intelligent Secure Systems, KFUPM, Dhahran 31261, Saudi Arabia*

[5]*Center for Advanced Material, King Fahd University of Petroleum and Minerals, Dhahran 31261, Saudi Arabia*

*Corresponding author: sens@iiti.ac.in*

**Abstract:**

A new study explores the distinct roles of spontaneous polarization and piezoelectric polarization in piezo-phototronic coupling. This investigation focuses on differences in photocatalytic and piezo-photocatalytic performance using sodium bismuth titanate (NBT), a key ferroelectric material. The research aims to identify which type of polarization has a greater influence on piezo-phototronic effects. A theoretical assessment complements the experimental findings, providing additional insights. This study explores the enhanced piezo-phototronic performance of electrospun nanofibers compared to sol-gel particles under different illumination conditions (11W UV, 250W UV, and natural sunlight). Electrospun nanofibers exhibited a rate constant ($k$) improvement of 2.5 to 3.75 times, whereas sol-gel particles showed only 1.3 to 1.4 times higher performance when ultrasonication was added to photocatalysis. Analysis using first-principle methods revealed that nanofibers had an elastic modulus ($C_{33}$) about 2.15 times lower than sol-gel particles, indicating greater flexibility. The elongation of lattice along z-axis in the case of nanofibers reduced the covalency in the Bi-O and Ti-O bonds. These structural differences led to reduced spontaneous polarization and piezoelectric stress coefficients ($e_{31}$ & $e_{33}$). Despite having lower piezoelectric stress coefficients, higher flexibility in nanofibers led to a higher piezoelectric strain coefficient, 2.66 and 1.97 times greater than sol-gel particles, respectively. This improved the piezo-phototronic coupling for nanofibers.

Recent advancements have showcased the effective use of piezo potential and polarization charges as a strategic approach to enhancing the transport behaviour of electrons and holes within both the lattice and surface of photocatalysts [1–3]. This development has spurred rapid progress in various fields, including transistors, nanogenerators, LEDs, and solar cells [4]. A contemporary approach is to integrate piezoelectricity, photoexcitation, and semiconductor properties also known as piezo-phototronic coupling to introduce mechanical stimulation-induced polarization to generate a built-in electric field [5]. However, an experimental result explaining the mechanism through theoretical assessment is necessary to understand such phenomena.

Such piezo-phototronic effect is observed in non-centrosymmetric semiconductors. Several ferroelectric/piezoelectric simple oxides and perovskite oxides like ZnO, $SrTiO_3$, KNN, NBT, $BaTiO_3$, etc. are utilized in piezo-phototronic coupling [6–10]. Piezoelectric nanomaterials like $BaTiO_3$ nanowires display enhanced piezocatalytic activity compared to $BaTiO_3$ nanoparticles due to their higher piezoelectric potential, enabled by their large bending compliance and flexibility, allowing efficient conversion of mechanical strain into charge carriers even at minimal strain levels [11]. Hence, piezocatalysis/piezo-photocatalysis is a surface morphology dependent phenomenon.

Among various piezoelectric perovskite oxides, sodium bismuth titanate (NBT) ceramic is a potential material due to its high piezoelectric coefficient ($d_{33}$~70 pC/N), high remnant polarization ($P_r$~38μC/cm$^2$), and chemical stability [12–14]. Also, it possesses a high electrical conductivity (~10$^{-5}$ S/m) among other perovskites which is required for enhancement of the catalytic performance [8]. A work by Zhao et. al used hydrothermal synthesized NBT nanospheres to degrade RhB dye (k~0.061 min$^{-1}$) in the presence of a Xe-lamp [8]. Zhou et. al degraded various cationic dyes using piezo-phototronic effect of NBT nanosphere (hydrothermal method) and nanorods (molten salt method) using a Xe-lamp where nanorods (k~0.075 min$^{-1}$ for MB) outperformed the nanospheres [15]. It was reported that nanorods possess a higher $d_{33}^*$ ~200 pm/V and simulated (Finite element method) piezopotential of -1.49 V than the nanospheres ($d_{33}^*$ ~100 pm/V, -0.27 V). A study by Liu et. al reported piezocatalytic degradation of MB dye (~54.2% in 150 min) using electrospun 0.93($Bi_{1/2}Na_{1/2}$)$TiO_3$-0.07$BaTiO_3$ nanofibers [16]. Ji et al. utilized topochemical molten salt synthesis route grown thin sheets of $Na_{0.5}Bi_{0.5}TiO_3$ having strong ferroelectric polarization to study the photocatalytic performances [17]. A piezoelectric constant ($d_{33}$) of ~15 pm $V^{-1}$ was reported by Ghasemian et. al for the hydrothermally grown NBT nanofibers [18]. Apart from considering only the piezo response of NBT nanostructures, oxygen vacancy also plays an important role in piezocatalytic activity.

A first principle assisted study by Liu et. al reported that NBT with low O-vacancy exhibit robust piezoelectric properties [19]. This facilitates the generation of a large piezoelectric potential and accelerated charge transfer. However, the weak capacity to adsorb $O_2$ and $OH^-$, low electron concentration, and poor charge transfer and transport performance ultimately limit their piezocatalytic activity. Conversely high O-vacancy demonstrate improved adsorption capabilities for $O_2$ and $OH^-$, higher electron concentration, and enhanced charge transfer capacity. The accompanying reduction in $d_{33}$ and weakening of the piezoelectric potential impair their piezocatalytic performance, showcasing the double-edged-sword role of O-vacancy in piezocatalysis. Hence, the piezo-photocatalytic performance of NBT nanostructures is significantly dependent on the chosen preparation technology, which governs key factors including morphology, size, oxygen vacancy, piezoelectric properties, and photoexcited response.

Despite many available literatures on various nanostructured NBT for piezo-phototronic coupling applications, a clear understanding of how nanofibers elevate the piezopotential in the NBT unit cell is still missing. A clear distinction between the ferroelectric spontaneous polarization and piezoelectric polarization needs special attention to understand which type of polarization exactly influences the piezo-phototronic coupling [20,21]. To study the differences between the spontaneous and piezoelectric polarization, a novel investigation is being performed to find the differences between a photocatalytic and piezo-photocatalytic performances of a prominent ferroelectric, as in NBT. A theoretical assessment is also performed to substantiate the findings. To examine the importance of piezoelectric contribution two different samples with different morphologies is being discussed.

Na$_{0.5}$Bi$_{0.5}$TiO$_3$ were synthesized using two different synthesis routes: sol gel and electrospinning. These particles are referred to as: sol gel particles (SGP) and nanofibers (NF). Various characterization techniques like XRD, Raman spectroscopy, SEM/TEM, DRS, and XPS were performed to distinguish the properties of SGP and NF. As an extra information for the photocatalysis, two light sources (low cost 11 W UV and high cost 250 W UV lamps) were used to compare the cost efficiency of the process. Also, natural sunlight was used to further reduce the experimental cost to show the utilization of NF as a catalyst. An attempt is made to understand the enhanced piezo-photocatalytic performance of the NF in terms of the structural parameters (c/a, tilt angle, and octahedral strain), phonon modes (Ti-O and TiO$_6$ vibrations), and piezoelectric coefficient ($d_{33}$). A theoretical approach is utilized to envisage the differences in the SGP and NF in terms of the lattice strain, electron localization, and the Born effective charges. Hence, a clear understanding of the improved flexibility, piezoelectric constant ($d_{31}$ and $d_{33}$), and piezoelectric polarization of the NF as compared to the SGP is thoroughly explained.

**Results and Discussion:**

The X-ray diffraction pattern revealed the presence of a single phase in both the SGP and NF samples. The broadening of the XRD pattern for the NFs revealed a smaller particle size than the SGP. The Rietveld refinement confirms the samples are in the rhombohedral ($R$3c) phase [Fig. S1(a-b)]. The lattice parameters a=b and c for NF (5.5146(24) Å, 13.5059(32) Å) showed a higher value than the SGP (5.5066(20) Å, 13.4574(9) Å) sample. c/a ratio represents the lattice distortion along the c-axis which influences the piezoelectric properties of a material. A higher c/a and rhombohedral lattice distortion ($\eta_R = \frac{c}{\sqrt{6}a} - 1$) was observed for the NF [Table 1]. Octahedral tilt ($\omega = 4 \times \sqrt{3} \times e$) is an important parameter to understand ferroelectric perovskites that are derived from the atomic positions using the Megaw Darlington approach [22]. The tilt angle is higher for the NF (5.26°) than the SGP (4.06°) sample. "ζ" is the octahedral strain, given by: $\zeta = cos(\omega)\left(\frac{C_H}{\sqrt{6}a_H}\right) - 1$. The parameter 1+ζ indicates the elongation and flattening of the octahedra along the triad axis. 1+ζ is also higher for the NF (0.99567) than the SGP (0.99521). A higher c/a and octahedral strain generate more distortion which enhances the piezoelectric performance in NBT-based materials [12]. Hence based on structural analysis, it can be inferred that the nanofibers can be an advantage for piezoelectric applications as compared to the SGP samples.

**Table 1** Tabulation of the lattice parameters, volume, c/a, rhombohedral distortion, tilt angle, and octahedral distortion parameters derived from the CIF file after Rietveld refinement of the sol-gel prepared (SGP) and electrospun nanofiber (NF) NBT.

| Sample | Lattice Parameters | | c/a | Volume (V) (Å$^3$) | Rhombohedral Distortion ($\eta_R$) | Tilt angle ($\omega$) | Octahedral Distortion (1+$\varsigma$) |
|---|---|---|---|---|---|---|---|
| | a=b (Å) | c (Å) | | | | | |
| SGP | 5.5066(92) | 13.4574(9) | 2.4438 | 353.40(3) | -0.0023 | 4.06º | 0.9952 |
| NF | 5.5146(24) | 13.5059(3) | 2.4491 | 355.70(3) | -0.0001 | 5.26º | 0.9957 |

According to group theory, rhombohedral (*R*3c) NBT should possess 13 Raman active modes ($\Gamma_{R3c}$= 4A$_1$+ 9E). The crystallographic sites 6(a) (Na/Bi/Ti positions) correspond to four Raman (A$_1$+2E) whereas crystallographic sites 18(b) (O-atom position) in hexagonal settings correspond to nine Raman (3A$_1$+6E) active modes [23]. However, the experimental Raman spectra for NFs, in the range of 50-1000 cm$^{-1}$, revealed only seven major modes at 66 cm$^{-1}$ (A$_1$), 122 cm$^{-1}$(E), 279 cm$^{-1}$(A$_1$), 531 cm$^{-1}$(E), 604 cm$^{-1}$(E), 774 cm$^{-1}$ (A$_1$), and 857 cm$^{-1}$ [Fig. S1(c)]. The Raman shifts for the SGP was observed to be at 66 cm$^{-1}$ (A$_1$), 128 cm$^{-1}$(E), 279 cm$^{-1}$(A$_1$), 531 cm$^{-1}$(E), 604 cm$^{-1}$(E), 774 cm$^{-1}$ (A$_1$), and 857 cm$^{-1}$ [Fig. S1(d)]. The FWHM of the NF and SGP were found to be similar. The intensity of the nanofibers reduced than the SGP sample. The changes in the Raman spectra can be attributed to the variations in the synthesis route. These affect the particle size and strain. The local structural changes and the associated strain affect the phonons, which can be detected using Raman spectroscopy.

"E" mode for NF was observed at 122 cm$^{-1}$, while for the SGP it was at 128 cm$^{-1}$. The A-O bond length was observed to increase from 2.7576 Å (SGP) to 2.7625 Å (NF). This hints at the decrease of the bond strength, which may be the reason for the redshift. However, the changes in the Raman shift were negligible for the other modes.

A prominent observation was in the change in the intensity ratio of the A$_1$ mode at 66 cm$^{-1}$ to the A$_1$ mode at 279 cm$^{-1}$ for the NF and SGP. The ratio was found to be 1.13 for NF and 1.25 for SGP. Similarly, the intensity ratio of the A$_1$ (66 cm$^{-1}$) mode to the E mode at 531 cm$^{-1}$ also showed a difference between the NF (0.84) and the SGP (0.90). Hence, these intensity ratios were observed to decrease for both cases, as in the case of NF [Fig. 1(d)].

For the NFs, the FWHM changes significantly for two modes. The E mode (~604 cm$^{-1}$, TiO$_6$ octahedral vibration), showed an increase in FWHM from ~68 cm$^{-1}$ (NF) to 70 cm$^{-1}$ (SGP). On the other hand, the FWHM of the A$_1$ breathing mode (~774 cm$^{-1}$) decreased from ~94 cm$^{-1}$ (NF) to 92 cm$^{-1}$ (SGP). The broadening of the E mode could be associated with the strain induced in the lattice, which is also responsible for the reduction of the crystallite size of the NF. On the other hand, the nominal sharpening of the breathing A$_1$ mode can be due to a decrease in octahedral tilt (ω) in SGP compared to the NF.

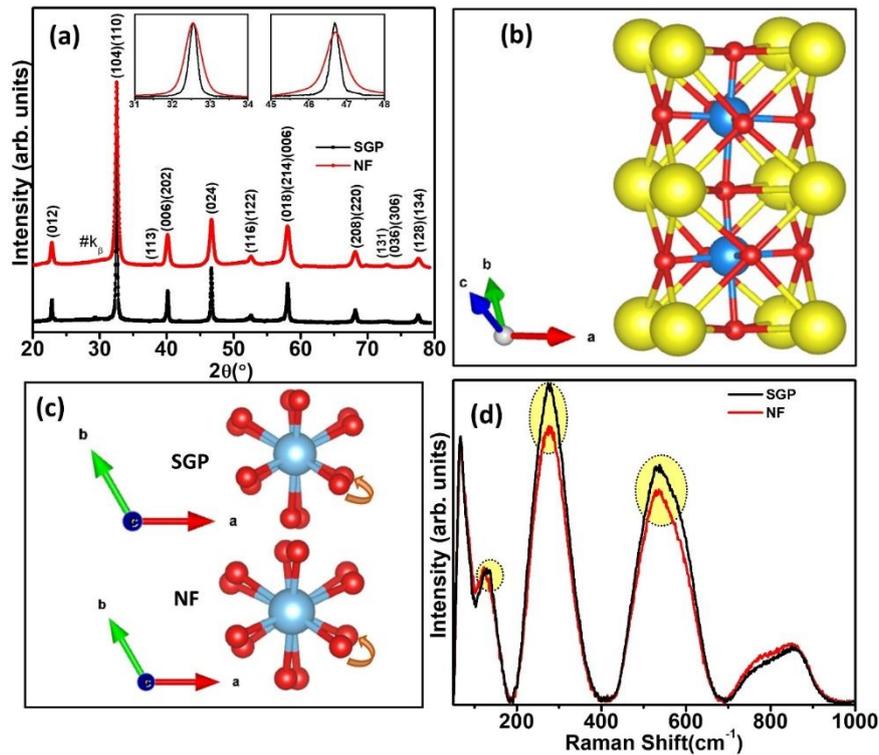

**Figure 1** (a) X-Ray diffraction pattern for sol-gel prepared NBT (SGP) and electrospun nanofiber (NF) (b) perovskite structure diagram for electrospun nanofiber (NF) (c) tilt angle of alternate TiO$_6$ octahedra for sol-gel prepared NBT (SGP) and electrospun nanofiber (NF) (d) Raman spectra for sol-gel prepared (SGP) and electrospun nanofiber (NF). [yellow atoms→A-site occupying atoms (Na/Bi), Blue atoms→ Ti, red atoms→ O].

The FESEM studies enabled the understanding of the formation of the nano crystallites in case of the SGP and the NF. The morphology of the SGP samples revealed agglomerated particles (~2.068 μm) of almost spherical morphology. However, these particles appear to be constituted of sub-100 nm nano crystallites. On the other hand, the electro-spun sample showed elongated tubular nanofibers of length 4-5 um and with diameters of ~273 nm. However, these nanofibers also seem to be constituted of sub-100 nm nano crystallites. The average nano-crystallite size of the SGP sample appeared larger than the NF. Note that the tubular nanofibers were derived from as-deposited fibers of the diameter of ~894 nm which were further annealed at 700 ºC to obtain the final phase-formed nanofiber of a smaller diameter of ~273 nm. The heating at 700 ºC, also shortened the length of the fiber from several micrometers to a few micrometres. Note that the as-deposited fibers appeared to be a continuous thread-like structure that became an array of nanoparticles upon heating due to surface/grain boundary diffusion and local evaporation-condensation processes [24].

To understand the structure and nano crystallite morphology Transmission Electron Microscopy (TEM) was performed. Both samples revealed nano-sized particles, of size ~70nm for SGP and ~40nm for NF [Fig.3(a)&(d)]. Selected Area Electron Diffraction (SAED) images revealed at least one circle constituted of several bright spots. However, one can analyse the TEM results to be constituted of complex diffraction patterns from individual crystallites in different orientations resulting in as many as seven concentric rings [Fig.3(b-e)]. The d-spacing obtained from these concentric circles matches with the d-spacings of the XRD results. High-Resolution Transmission Electron Microscopy (HRTEM) unveiled nuanced phase contrast images, illuminating distinctive grain orientations corresponding to (110) and (012) crystal planes, delineated by d-spacing dimensions of 0.27 nm and 0.35 nm, respectively [Fig. 3(c-f), Fig. S2&S3]. Moreover, elemental analysis elucidated the presence of only Sodium (Na), Bismuth (Bi), Titanium (Ti), and Oxygen (O) constituents within both the SGP and NF, thereby ensuring the phase purity of the materials [Fig. S4].

The $N_2$ adsorption-desorption curve obtained from the BET measurement revealed a higher specific surface area of ~10.573 $m^2$/g for the NF calculated from the adsorption curve [Fig. S6]. The pore volume and the average pore diameter were found to be 0.075 cc/g and 17.58Å. The SGP sample did not reveal consistent results may be due to the less porous structure.

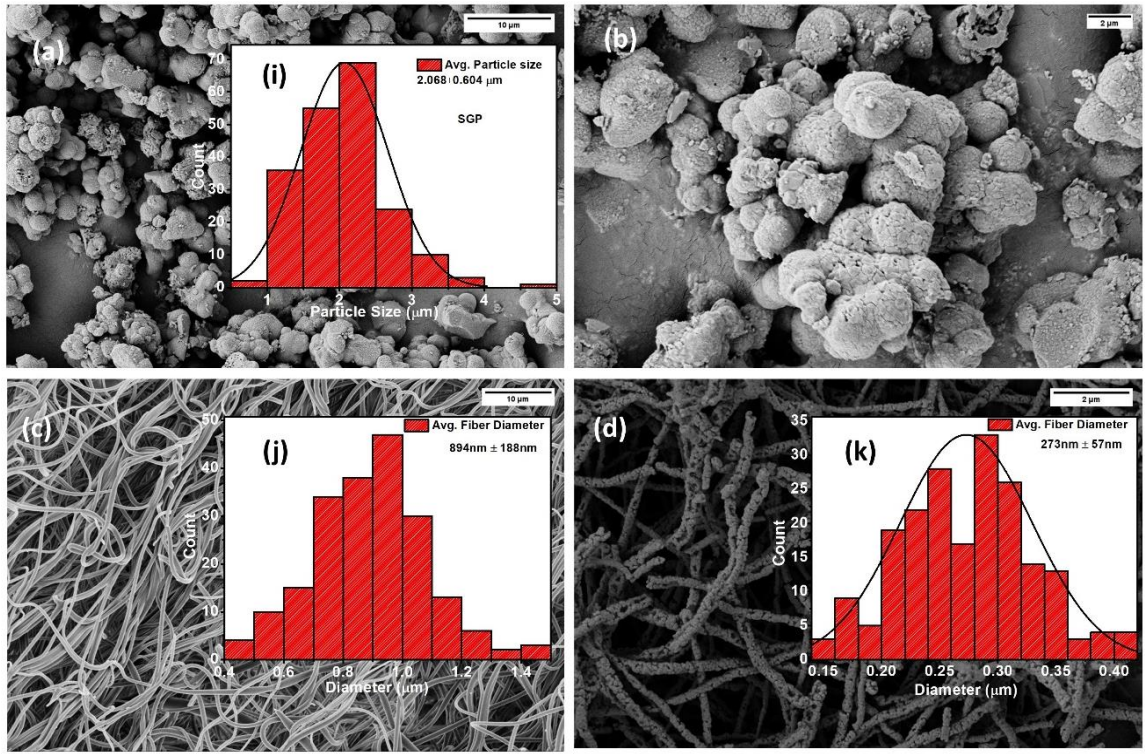

**Figure 2** FESEM images of (a) sol-gel prepared SGP NBT (b) sol-gel prepared SGP NBT showing agglomerated spherical morphology (c) electrospun NBT nanofibers (NF) after drying at 120ºC (d) electrospun NBT nanofibers (NF) annealed at 700ºC. [(i) particle size distribution of agglomerated microspheres of sol-gel NBT (SGP) (j) nanofiber diameter distribution of 120ºC dried electrospun NBT (NF) (k) nanofiber diameter distribution of 700ºC annealed and phase formed electrospun NBT (NF)].

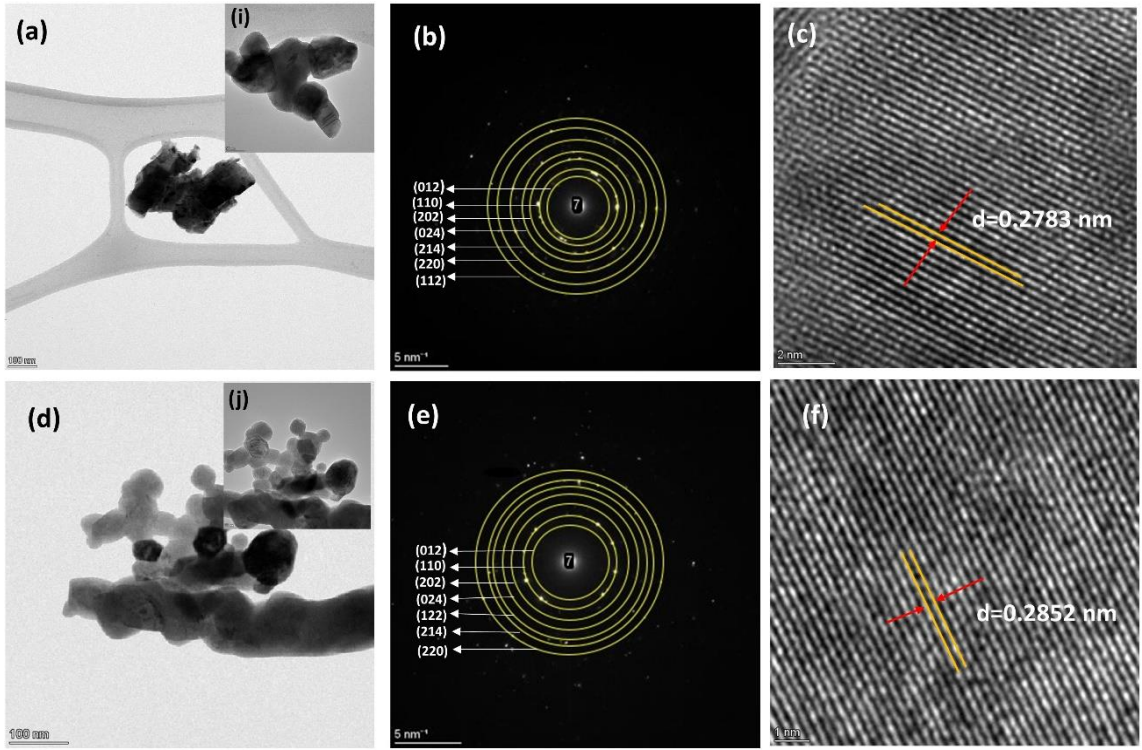

**Figure 3** (a) TEM image of sol-gel prepared NBT (SGP) (b) SAED pattern of SGP confirming *R*3c phase (c) HRTEM image of SGP showing d-spacing for (110) plane (d) TEM image of electrospun NBT nanofiber (NF) (b) SAED pattern of NF confirming *R*3c phase (c) HRTEM image of NF showing d-spacing for (110) plane. [TEM image of NBT at 50nm scale (i) sol-gel prepared SGP (j) electrospun NBT nanofiber (NF)].

The modified Kubelka Munk equation given by the relation: $F(R_\infty) \propto \frac{(h\nu - E_g)^{\frac{1}{n}}}{h\nu}$, $(F(R_\infty)h\nu)^n = A(h\nu - E_g)$ was used to calculate the band gap of SGP and NF [25,26]. For the direct band gap transition, "n" is considered to be 2 and for the indirect transition, ½ is used. A plot between $(F(R).h\nu)$ n vs "energy ($h\nu$)" known as the Tauc plot was used. It generated a straight line with an x-axis intercept equivalent to the bandgap ($E_g$) for both direct and indirect cases. In literature, NBT-based samples have been reported to have both direct and indirect band gaps [27]. In this case, NF revealed a higher direct band gap ($E_{gd}$=3.34 eV) than the indirect band gap ($E_{gi}$=3.25 eV). Similarly, for the SGP samples, the $E_{gd}$ was found to be 3.17 eV while $E_{gi}$ was 3.08 eV. Note that both values are lesser for the SGP than the NF. To understand this behaviour a theoretical calculation is essential which will be discussed in the following section.

Disorders and strain in lattice can result in tailing states near the band edges resulting in an exponential increase of $F(R)$, with energy ($h\nu$). These tailing states are known as the Urbach tail and can be related to the Urbach energy ($E_U$) as: $F(R) = (F(R))_0 exp(\frac{h\nu - E_g}{E_U})$, where, $F(R)_0$ is a material constant [28]. The above can be modified by taking logarithms on both sides and written as: $ln(F(R)) = ln(F(R))_0 + \frac{h\nu - E_g}{E_U}$. Hence, the inverse of the slope obtained from linear fitting of the plot between ln($F(R)$) vs "$h\nu$" gives $E_U$. Urbach energy was observed to be lesser for the NF (75 meV) in comparison with the SGP (119 meV). This indicates a decrease in defect-associated strains in nanofibers, which could be due to the decrease in oxygen vacancy from 26% for SGP to 16% for NF. This proposition is further supported by XPS analysis (Fig. S5).

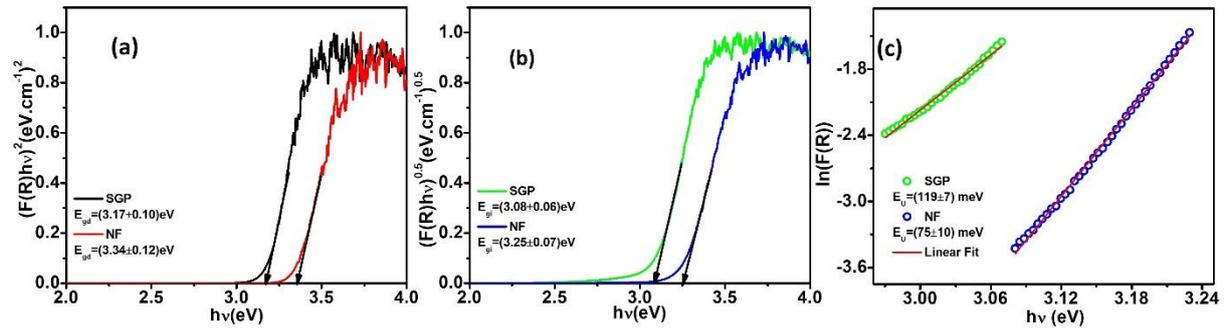

**Figure 4** (a) Direct band gap (b) indirect band gap (c) Urbach energy for SGP and NF.

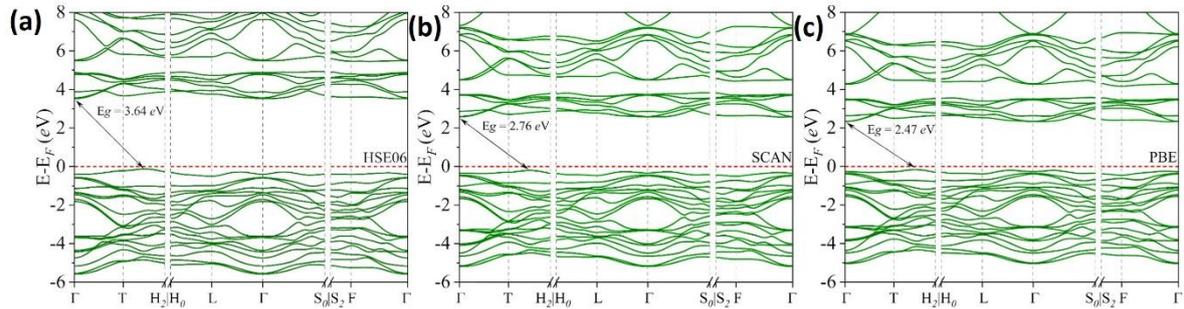

**Figure 5** Band-structure of NBT (SGP) calculated using (a) HSE06 (b) SCAN, and (c) PBE-DFT functionals. The Fermi level is set to 0 $e$V, keeping the equilibrium geometry fixed, obtained from GGA-PBE functionals.

The spin-polarized density functional calculations using VASP revealed the theoretical electronic structure of the ideal NBT-SGP structure at its equilibrium volume. Employing GGA approximations, a band gap of 2.47 eV was calculated, which was noticeably lower than the experimentally observed indirect band gap of 3.08 eV. This discrepancy underscores the inherent limitations of GGA methods in accurately describing the electronic structure. This is likely due to the lack of exact electronic correlations. A single DFT functional that can perfectly describe the electronic structure of materials is impossible. Hence, a few more DFT functionals, such as Hybrid (HSE06) and metaGGA (SCAN) were considered. A k-resolved DOS is plotted along the high-symmetry BZ path for all three approximations (Fig. 5) to understand the hybridization within the vicinity of the electronic gap. The calculated band gaps using the HSE06 (3.64 $e$V) and SCAN (2.76 $e$V) functionals are shown in Fig. 5(a) & (b).

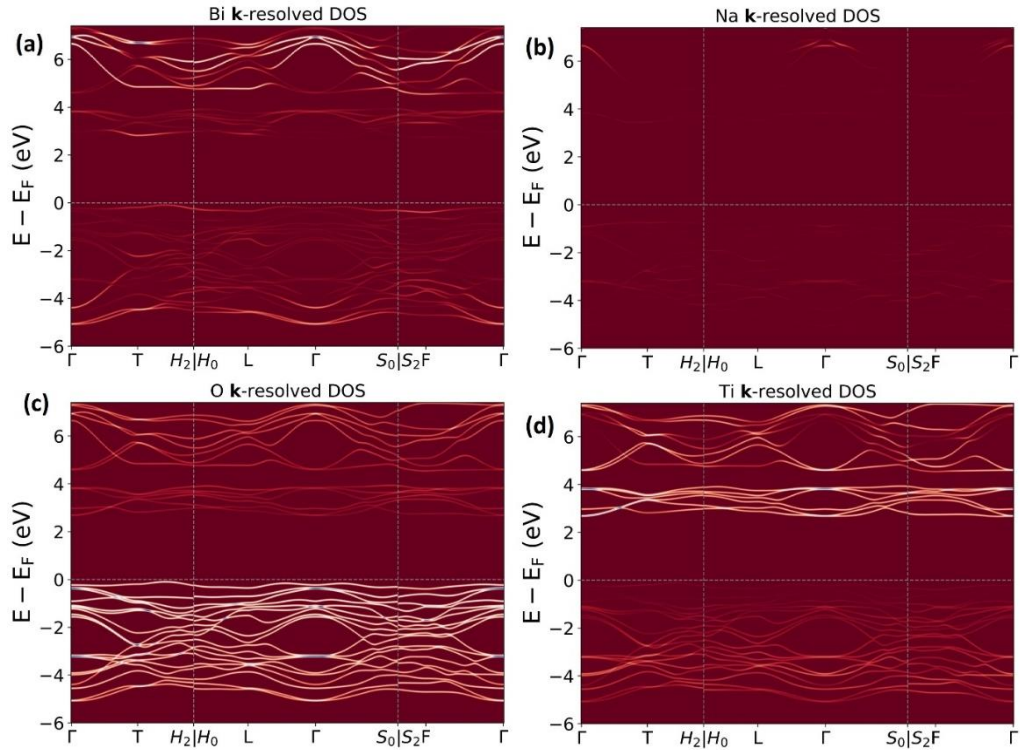

**Figure 6** An atom-resolved k resolved (DOS) projected along the high-symmetry lines calculated using the SCAN DFT functionals for (a) Bi-atom (b) Na-atom (c) O-atom (d) Ti-atom. [The conduction band is majorly populated with Ti-states, whereas the valence band is with O-states].

Amongst the three approximations, the SCAN DFT functional is in much closer agreement with experimental observations. The conduction band minima (CBM) lie at the $\Gamma$-point, and the valence band maxima (VBM) lies between the T-H2. A k-resolved DOS is plotted along the high-symmetry BZ path for each element (Fig. 6) to understand the hybridization within the vicinity of the electronic gap. The Ti-states majorly contribute to the CBM, with some partial occupation of Bi and O-states. In the case of VBM, it is mainly occupied by the O-states with some partial Bi occupation. The Fermi level is observed very close to the valence band i.e., the O-states. Na-states are highly stabilized at the Fermi level.

With the evolution of the particle's size and shape, an internal strain is generated. This strain significantly modifies the lattice structure and thereby modifies the electronic structure. Several scenarios of lattice strain were considered, e.g., isotropic, lateral, and uniaxial strain, keeping the volumetric strain constant for each case. The electronic gap's nature in NBT nanofibers was simulated using SCAN DFT functionals, considering these potential scenarios.

The calculated bandgaps are 2.82 $eV$ (expansion) and 2.57 $eV$ (compression) for Isotropic strain. Similarly for lateral strain, these values were 2.68 eV (expansion) and 2.55 eV (compression), while for uniaxial strain the values were 2.99 eV (expansion) and 2.03 $e$V (compression) [Fig. S7]. The difference between the SGP and NF band gaps can be observed experimentally to be ~0.17 eV for both direct and indirect band gap. Theoretically, all the calculations show an indirect band gap. The indirect $E_g$ from the SCAN calculations was 2.76 eV for the NBT SGP sample. The differences of this $E_{gi}$ were calculated with all the six possible cases of a strained NBT lattice. From these differences, the uniaxial expansion revealed a difference of ~0.23 eV. Hence, it appears that the uniaxial expansion is the most feasible strain scenario in these nanofibers (NF). Following the observations on electronic structure and experimental observations, only the DFT-optimized equilibrium and uniaxial expansion structure were considered for further analysis of piezoelectric properties.

**Photocatalysis /Piezo-phototronics effect:**

To understand the effect on the piezoelectric properties in the presence of light a simple process has been investigated in this work in terms of understanding the changes in the photodegradation process in the presence of light and external mechanical vibrations. As a source of mechanical vibrations, an ultrasonication process has been adopted. The degradation of methylene blue (MB) dye (5 ppm) was investigated in the presence of a 40kHz ultrasonic bath while irradiating with photons of two different types of artificial sources: (1) LPMV 11W UV light and (2) MPMV 250W UV light.

Four different concentrations of NF were experimented with: 0.5g/L, 1g/L, 1.25g/L, and 1.5g/L. The degradation increased with the NF concentration from 65% (0.5g/L) to 68% (1g/L), and a maximum of ~71% for 1.25g/L under UV (11W) [Fig. S8]. Further increasing the concentration reduced the degradation rate. Increasing the catalyst amount in heterogeneous photocatalysis leads to intensified photodegradation due to elevated collisions between reactants and increased formation of OH• radicals on the photocatalyst surface. However, beyond a threshold catalyst quantity, turbidity in the solution impedes the reaction by obstructing the necessary radiation, resulting in a decline in degradation percentage [29]. Hence, 1.25g/L concentration was selected for further investigation to understand the piezo-photo coupling in SGP and NF samples.

A similar study was performed without the vibrations on the 1.25g/L SGP and nanofiber samples to understand the effect of the ultrasonic vibrations. The time-resolved absorbance spectra are shown in Fig. S8. With time, there was a steady decrease in MB concentrations. The degradation efficiencies in percentage were determined using the equation: $Degradtion \% = (\frac{C-C_0}{C_0}) \times 100$. Degradation was observed for all the concentrations for both samples revealing a universality of the catalytic activity of the NBT samples. However, NF degraded the MB solution at a higher rate for all irradiation types and intensities than the SGP one [Fig. 7]. The degradation for nanofibers was ~30% after 180 min illumination of the 11W UV light, while for SGP, it was ~23%. For the 250W illumination, the degradation after 180 min was ~41% for the nanofiber and ~36% for the SGP sample [Table 2]. The degradation increase under higher power illumination may be due to the available number of photons in the 250W source. However, considering the bandgap to be ~3.08 eV, such utilization of lesser energy light is less probable until one considers defect states in the samples. However, one must also consider that an intensity dependence is not a straightforward correlation, considering the difference in powers of the two light sources. This aspect thus instigates a further detailed intensity-dependent study of the two individual sources of the degradation mechanism. However, the scope of this work is to focus on the effect of sonication on the photodegradation process. Both nanofibers and SGP showed enhanced degradation under the simultaneous influence of light and ultrasonic vibration. The degradation for the NF samples was observed to be ~71% with the 11W and ~85% with 250W light illumination for 180 mins. The SGP revealed a much lesser degradation of ~26% for 11W and ~50% for, 250W for the same illumination time, i.e. 180 mins [Table 2]. Hence, the NF exhibit better photocatalytic activity than the SGP counterpart with and without the involvement of sonication. This could be related to the high specific surface are of NF.

Photocatalysis in semiconductors generally comprises three underlying steps [30,31]:

(1) photon absorption and generation of electron-hole pairs: light irradiation on the photocatalyst promotes the transfer of electrons from the valence band to the conduction band leaving holes in the valence band and thereby creating electron-hole pairs.

(2) separation of excited charges and migration to the surface from lattice: electron-hole pairs generated by light absorption are separated before they recombine. This separation is crucial for achieving efficient catalytic reactions because it allows the charges to participate in redox reactions at different sites of the photocatalyst material.

(3) involvement of the electrons and holes in the redox reactions at the surfaces.

The performance of any photocatalysts mostly depends upon the charge separation and transfer kinetics. During the photocatalysis, most electron-hole pairs undergo recombination, either during transit to the surface or upon reaching surface sites [32–34]. This recombination process results in the dissipation of harvested energy, that takes place in nonradiative-heat dissipation (capture) or radiative recombination (light). Such capturing or recombination of the photoinduced carriers in the lattice or surface significantly reduces the photocatalytic performance.

A piezoelectric material has an inbuilt dipole due to an application of external field or strain, giving rise to dipole polarization. The interaction between piezoelectric polarization and electronic transport leads to advantageous synergistic outcomes. Applying external strain to a piezoelectric material enables one to control charge transport, recombination, or separation of the charge centres. Hence, such an external strain-induced polarization in piezoelectric materials facilitates the process of manipulation of charge-carrier transport properties within the lattice [35,36]. Thereby, the movement of photoinduced charges i.e., electrons/holes can be manipulated restricting recombination by enhancing their physical separation. In the present study, the ultrasonic vibration generated the required mechanical vibrations which facilitate such a physical separation of the electron-hole pairs. From the experimental results, the NFs are observed to be more sensitive to mechanical vibrations than the SGPs. This may be due to the morphology of the nanofibers which can be more sensitive to the vibrations and thereby result in higher built-in polarization [15].

The kinetics of the dye degradation process can be expressed using the relation: $C = C_0 e^{-kt}$, where "$C$" is the concentration at time $t$, "$C_0$" is the initial concentration at $t$=0, and "$k$" is the first order rate constant. "$k$" was calculated from the linear fitting of ln($C_0/C$) and time ($t$) [Fig. 7(c-d)] for both the samples. The "$k$" value obtained for the photocatalytic effect in irradiated nanofiber was 0.0016 min$^{-1}$(11W) and 0.004 min$^{-1}$(250W) while for the irradiated SGP sample was 0.0014 min$^{-1}$(11W) and 0.002 min$^{-1}$(250W). Hence, the NBT nanofibers showed a higher "$k$" than the SGP sample. A higher "$k$" represents a better catalytic performance.

Similarly, for the piezo-phototronic effect "$k$" value obtained for the NF was 0.006 min$^{-1}$(11W) and 0.010 min$^{-1}$(250W) while for the SGP it was 0.0019 min$^{-1}$(11W) and 0.0028 min$^{-1}$(250W). Hence, it is observed that the "$k$" value has increased from 0.0016 min$^{-1}$ to 0.006 min$^{-1}$ for the 11W illumination with the application of sonication [Table 2]. This is an increase by

3.75 times establishing a strong piezo-phototronic effect in the NF for 11W illumination. Similarly, for the 250W illumination, NF exhibited 2.5 times enhancement for the sonication. These are noteworthy remarkable improvements in the piezo-phototronic catalytic performance of NBT nanofibers. Similarly, for the SGP samples the piezo-phototronic improvement were 1.35(11W) and 1.40 (250W) times. Hence, NF exhibit better improvement with the application of ultrasonic vibration than the SGP. One important information extracted from this study is achieving better performance of nanofibers even under low power(11W) illumination. This is a significant contribution to reduction of cost in performance of the functionality.

One more experiment was performed with intentions of reducing the cost of experiment using natural sunlight. A similar comparison was performed on both samples with natural sunlight illumination for 60 mins [Fig. S12]. The "$k$" value for the nanofibers were 0.015 min$^{-1}$ (photocatalysis) and 0.023 min$^{-1}$(piezo-photocatalysis). This is an improvement by 1.53 times using the ultrasonic vibration revealing a strong piezo-phototronic coupling in the nanofibers. On the other hand, the "$k$" value for the SGP sample were 0.004 min$^{-1}$ (photocatalysis) and 0.006 min$^{-1}$(piezo-photocatalysis). This is an improvement by 1.50 times using the ultrasonic vibration revealing a similar strong piezo-phototronic coupling in the SGP. This reveals a universal improvement any sort of illumination due to piezo-phototronic coupling in the nanofibers as well as SGP. Note that the "$k$" values of the SGP are quite lesser than the nanofibers in the case of natural sunlight. Hence, the piezo-phototronic catalysis in the presence of natural sunlight using nanofibers can be a prominent process of successful dye degradation in day time.

**Table 2** Degradation Percentage (%) and the rate constant (k) for different irradiation and vibration conditions for sol-gel particle (SGP) and nanofiber (NF). [L→ Only light, L+S→ Light + Sonication].

| **Degradation Percentages and Rate constants** | | **Samples** | |
|---|---|---|---|
| | | **SGP** | **NF** |
| **Degradation in 180 mins (%) [11W]** | L | 23 | 26 |
| | L+S | 30 | 71 |
| **Degradation in 180 mins (%) [250W]** | L | 36 | 51 |
| | L+S | 41 | 85 |
| **Degradation in 180 mins (%) [natural sunlight]** | L | 25 | 60 |
| | L+S | 32 | 75 |
| **Rate Constant (k) in min$^{-1}$ [$\times 10^{-3}$] [11W]** | L | 1.48 | 1.64 |
| | L+S | 1.93 | 6.65 |

| Rate Constant (k) in min$^{-1}$ [× $10^{-3}$] [250W] | L | 2.4 | 4 |
|---|---|---|---|
| | L+S | 2.8 | 10 |
| Rate Constant (k) in min$^{-1}$ [× $10^{-3}$] [natural sunlight] | L | 4.78 | 14.9 |
| | L+S | 6.41 | 23.2 |

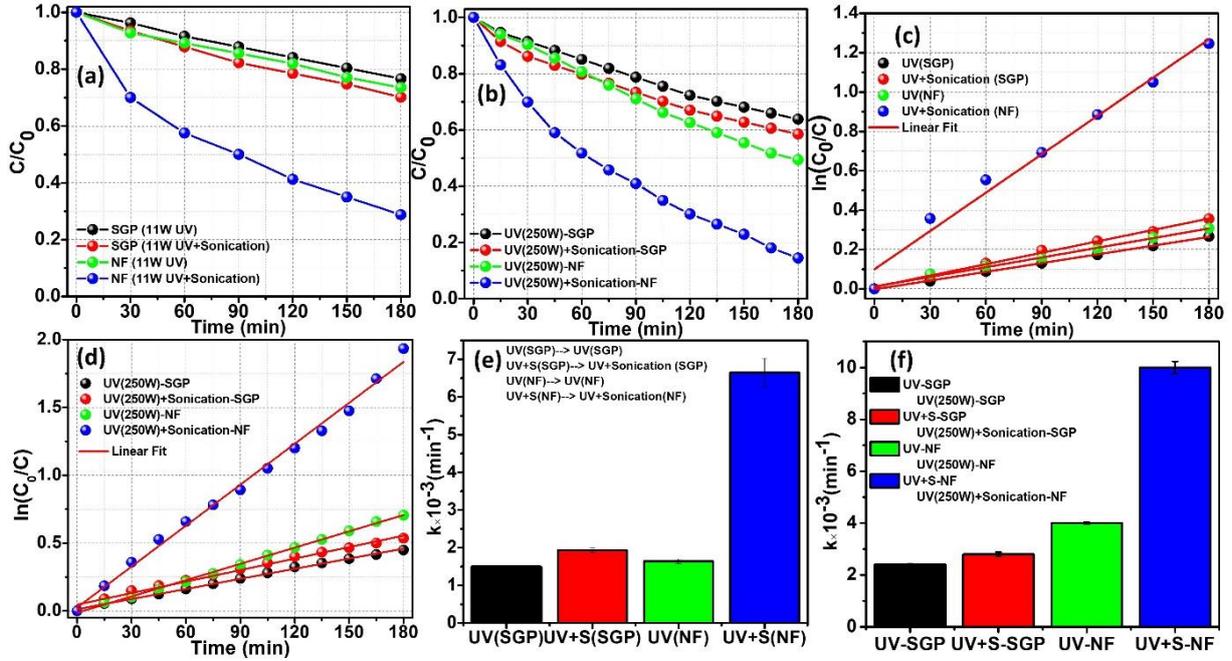

**Figure 7** Relative concentration change of MB dye with an increase in photon irradiation time for SGP and nanofiber under two conditions (a) 11W UV (b) 250W UV, Rate of change of MB dye concentration with time for both SGP and nanofiber under two conditions (c) 11W UV (d) 250W UV, Comparison of rate constant "k" for SGP and nanofiber for only photocatalysis and for piezo-photocatalysis under two conditions (e) 11W UV (f) 250W UV.

## *Ab Initio Simulation*

The induced polarization (along i, $P_i$) by a unit displacement (along j, $u_j$) of the sublattice of atoms gives rise to Born effective charge tensor ($Z^*_{ij}$): $Z^*_{ij} = \frac{\partial P_i}{\partial u_j}$ [37]. The off-centering of ions responsible for the ferroelectric effect can be understood by the Born effective charges [38,39]. The Born effective charges components along the diagonal ($Z^*_{xx}$, $Z^*_{yy}$, $Z^*_{zz}$) and off-diagonal ($Z_{xy}$, $Z_{yx}$) of the charge tensor for Na, Bi, Ti, and O atoms are given in the Table S2. As Na, Bi, and Ti atoms possess high local site symmetry, hence their effective charge tensors give rise to almost diagonal tensor having two independent components showing less anisotropy ($Z^*_{xx}=Z^*_{yy}$, $Z^*_{zz}$). Due to less symmetry in the O-atoms, the effective charge tensor is more anisotropic having all the independent components. It can be observed that the effective charges of Bi, Ti, and O atoms showed a significant deviation from the ideal ionic values (+3: Bi, +4: Ti, and -2:O). This represents the mixed ionic and covalent bonding in Bi-O and Ti-O due to the hybridization of Bi-6$p$ and Ti-3$d$ orbitals with the O-2$p$ orbitals for both SGP and NF [40].

For the Na atom, one independent component in the Born-effective diagonal tensor can be observed. However, there is a small off-diagonal antisymmetric component ($Z^*_{xy}=-Z^*_{yx}$) was observed for both SGP (0.047) and NF (0.054). This indicates that the Na atom is comparatively isotropic than other atoms. The effective charge density ($Z^*_{xx}=Z^*_{yy}$, $Z^*_{zz}$) value for both SGP (1.112, 1.126) and NF (1.140, 1.195) are close to the ideal ionic value i.e., +1. This was observed for both SGP and NF. The effective charges were observed to be nominally higher for the NF.

For the Bi-atom, two independent components were observed along the diagonal, one parallel to $z$-axis ($Z^*_{zz}$) and other two perpendicular to z-axis ($Z^*_{xx}=Z^*_{yy}$). The $Z^*_{xx}=Z^*_{yy}$ components are +5.364 for the SGP and +5.246 for the NF. Similarly, $Z^*_{zz}$ values are +3.772 for the SGP and +3.171 for the NF. These values are also higher than the ideal ionic value (+3). Also note that the nanofibers exhibited a lesser effective charge than the SGP for all the diagonal components. An off-diagonal component $Z^*_{xy}=-Z^*_{xy}$ was observed for both SGP (-0.651) and NF (-0.735). This makes Bi comparatively anisotropic than Na atom.

For the Ti1-atom (0, 0, $z$), two independent components were also observed. The $Z^*_{xx}=Z^*_{yy}$ components are +6.173 for the SGP and +5.243 for the NF. Similarly, $Z^*_{zz}$ values are +6.099 for the SGP and +4.614 for the NF. These values are significantly higher than the ideal ionic value of +4. It can be observed that the NF exhibited a significantly lesser effective charge than the SGP for Ti1 atom.

For the Ti2-atom (1/3, 2/3, $z$), two independent components were observed along the diagonal of the tensor, one parallel to $z$-axis ($Z^*_{zz}$) and other two perpendicular to $z$-axis ($Z^*_{xx}=Z^*_{yy}$). The $Z^*_{xx}=Z^*_{yy}$ components are +7.403 for the SGP and +6.812 for the NF. Similarly, $Z^*_{zz}$ values are +6.099 for the SGP and +4.614 for the NF. These values are significantly higher than ideal ionic value of +4. It can be observed that the nanofibers exhibited quite a lesser effective charge than the SGP for all the diagonal components for Ti2 atom. An off-diagonal component $Z^*_{xy}=-Z^*_{yx}$ was observed for both SGP (0.845) and NF (0.875). This makes Ti2 comparatively anisotropic than Na atom. The dissimilarity in $Z^*$ of the two Ti ions is due to hetero-ferroactive disorder at both the A and B sites in NBT which gives rise to the relaxor behavior of NBT [40].

The eigenvalues of the effective charge tensor vary in the range of -1.387 to -5.306 for SGP and -1.121 to -4.887 for NF. Such a significant deviation from the ideal ionic value of -2 indicates that the O atoms favour covalent bonding.

Compared to Bi and Ti, a nominal deviation of effective charge in Na atom reflects that the Na-O bonds in NBT are mostly ionic in nature. Considering a significantly higher Born effective charge of the Bi and Ti atoms for the SGP sample compared to the NF, it can be concluded that the SGP showed a higher covalency than the NF. This nature of the Bi-O bonds and Ti-O bonds for SGP and NF can be visualized in the Fig. 8. In this Figure, due to the presence of higher covalency for the SGP sample, the electrons sharing is stronger for the Ti-O and Bi-O electronic cloud which can be understood from the higher overlapped regions. The weaker hybridization (weak covalency and high ionicity) of the Bi-O and Ti-O indicate towards a weaker bond strength for the NF than the SGP. This was also observed in the higher average A-O and B-O bond length or lesser bond strength for the nanofibers (A-$O_{avg}$: 2.7625 Å, B-$O_{avg}$: 1.9551 Å) than the SGP (A-O: 2.7576 Å, B-O: 1.9496 Å) from the XRD study. Hence, the elongation of NBT lattice along the *z*-axis (NF) weakened the covalency. The lesser covalency in the nanofibers is responsible for obtaining a lesser spontaneous polarization than the SGP NBT [41–43]. It can be inferred that the experimentally observed lesser intense Raman spectra for the NF than the SGP for the Ti-O and $TiO_6$ phonon modes are the consequence of the decreased polarizability of the Ti atoms [44].

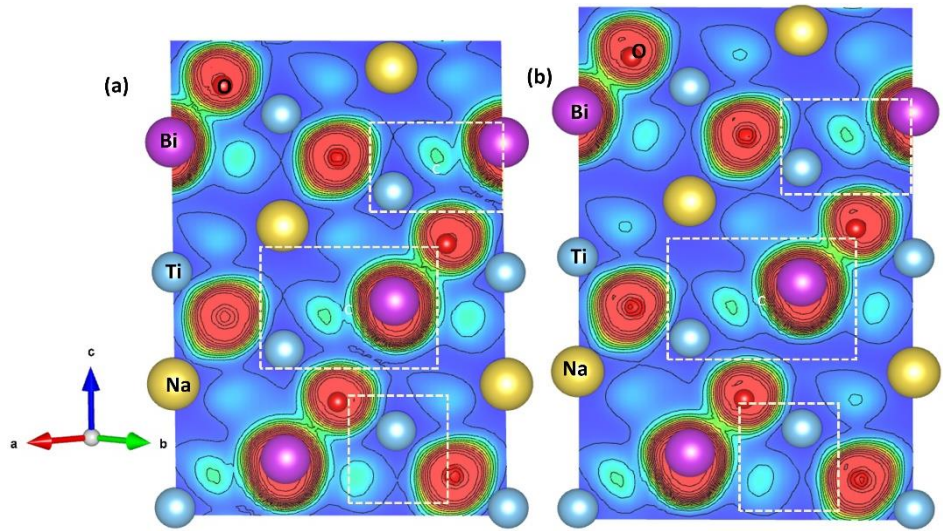

**Figure 8** Electron Localization function (ELF) plot for NBT in the (110) plane for (a) SGP (b) uniaxially elongated NBT (NF).

**Piezoelectric properties**

To understand the better piezo-phototronic coupling in nanofibers (NF) compared to the SGP, various parameters like elastic constants, piezoelectric stress coefficients, and piezoelectric strain coefficients were calculated. Various mechanical properties like stiffness, ductility, stability, and type of bonds can be determined from the elastic stiffness constants ($C_{ijkl}$) [45]. NBT structure belongs to a trigonal space group (161) with a point group of '$3m$'. The stiffness tensor ($C_{ijkl}$) was calculated from: $C_{ijkl} = \frac{\partial^2 E}{\partial \delta_{ij} \partial \delta_{kl}}$, where, $E$ is the total energy, $\delta_{ij}$ is the stress tensor, and $\delta_{kl}$ is the strain tensor [46]. The calculated $C_{mn}$ (i, j, k, l=1,2,3; m, n=1 to 6) values are mentioned in Table 3.

**Table 3**: Elastic stiffness constants for SGP and for z-elongated structure (NF) of NBT (in GPa).

| Sample | $C_{11}$ | $C_{12}$ | $C_{13}$ | $C_{14}$ | $C_{33}$ | $C_{44}$ | $C_{66}$ |
|---|---|---|---|---|---|---|---|
| SGP | 229.5 | 97.03 | 48.26 | -22.15 | 153.54 | 27.54 | 66.23 |
| NF | 152.2 | 54.25 | -2.20 | 22.62 | 71.27 | -43.22 | 48.96 |

The elastic stiffness $C_{11}$ and $C_{33}$ indicates the compressive stiffness (compressive stress with respect to the strain in same axis) along the [100] and [001] directions respectively. The constants $C_{44}$ and $C_{66}$ describes the shear stiffness in a plane (stress with respect to the strain across a face) [47]. The $C_{12}$, $C_{13}$, and $C_{14}$ represents the modulus for dilation on compression (axial stress with respect to the strain which is perpendicular to the axis). The unidirectional elastic stiffness of NF shows a lesser value (152.2, 71.27) than the SGP (229.5, 153.54). Hence, the uniaxial stiffness of the SGP is about ~1.50 ($C_{11}$) and ~2.15 ($C_{33}$) times larger than the NF. The larger the stiffness constant, the less flexible the materials are. Hence, the nanofibers (NF) are more uniaxially flexible than the SGP. It seems that the flexibility of the materials is advantageous to the piezo-phototronic coupling effect [48].

The lack of inversion symmetry in the NBT structure results in intriguing piezoelectric properties. The piezoelectric stress coefficient ($e_{ij}$) is the total stress due to two contributions, namely the ionic and electronic [Fig. S13-16]. The ionic contribution represents the additional internal relaxation of the relative atomic coordinates that would be induced by the strain. The electronic contribution in the structure comes from the modifications in the local fields and are dependent on external factors. In these samples, for both nanofibers and SGP, the ionic contribution dominates. According to the crystal symmetry, four piezoelectric stress coefficients are important, $e_{31}$, $e_{33}$, $e_{15}$, and $e_{22}$. Amongst these four stress coefficients, $e_{22}$ is the uniaxial stress along [010] direction, while $e_{33}$ is the uniaxial stress along [001] direction, $e_{31}$ is the stress perpendicular to [001] direction, and $e_{15}$ is the shear stress.

The uniaxial stress coefficient $e_{22}$ reveals an opposite contribution from their corresponding ionic and electronic components, thereby reducing the effective $e_{22}$. Moreover, the effect of ionic is negative in the SGP (-1.90716) and positive in nanofiber (1.09094), while the effect of electronic is positive in the SGP (0.20572) and negative in nanofiber (-0.24227). Thereby the net contributions in $e_{22}$ are negative in the SGP but positive in nanofiber [Table4].

However, the internal strain (ionic) and the electronic contributions have the same sign for the other three stress coefficients. This enhances the net stress coefficients. Amongst these three $e_{31}$ and $e_{33}$ are positive for both the samples while $e_{15}$ are negative. The magnitudes of all the stress coefficients are lesser in nanofibers in comparison to the SGP which is the consequence of the calculated Born effective charges [49].

**Table 4** Calculated $e_{ij}$ and $d_{ij}$ matrix elements for the SGP and NF.

| Sample | $e_{ij}$ (matrix elements) [C/m$^2$] | | | | $d_{ij}$ (matrix elements) [pC/N] | | | |
|---|---|---|---|---|---|---|---|---|
| | $e_{31}$ | $e_{33}$ | $e_{15}$ | $e_{22}$ | $d_{31}$ | $d_{33}$ | $d_{15}$ | $d_{22}$ |
| SGP | 2.30407 | 3.83288 | -0.79979 | -1.70144 | 3.71 | 22.63 | 268.83 | 58.1 |
| NF | 1.94306 | 3.14289 | -0.16607 | 0.84867 | 9.89 | 44.71 | 257.94 | 52.24 |

The piezoelectric strain coefficient $d_{ij}$ can be directly related to the piezoelectric stress coefficient $e_{ik}$ and elastic compliance $S_{jk} = C_{jk}^{-1}$ by the relation; $d_{ij} = \sum_{k=1}^{6} e_{ik} S_{jk}$ [50]. Under the uniaxial strain along the z-axis, the important parameters are $d_{31}$ and $d_{33}$. The d-matrix for both samples are represented in 3D and 2D plots using MTEX to visualize the piezoelectric strain [Fig. 9] [51,52]. The piezoelectric strain matrix is represented by four lobes. Along the x-y plane there are three lobes on the same plane while the fourth is along the z-axis relative to these three lobes. However, this 3D representation of the d-matrix appears to be opposite in orientation for the SGP and NF [Fig. 9(a, d)]. The reason behind such an opposite orientation comes from the opposite signs of the shear strain ($d_{15}$). Note that the other three uniaxial strains are all positive for both samples. Also, it is noteworthy that the $d_{31}$ and $d_{33}$ values are significantly large for the NF than the SGP.

Analysing the 3D plot, one can obtain a 2D representation along the basal plane showing two halves of the spheres one along +z-axis and other along -z-axis. These are represented in Fig. 9(b,c & e,f). Note that the reversal of the lobes is well represented by the opposite signs (colors). In the case of SGP and NF, the piezoelectric strain is much intense in the x-y plane compared to that along the z-axis i.e., a higher strain is generated along the x-y plane. For the nanofibers the strain is intense in x-y plane as well as along the z-axis. The piezoelectric strain along the +z-axis for the SGP and NF is positive and the magnitude for the NF (44.71) is higher than the SGP (22.63). Along the -z-axis, similar magnitude of piezoelectric strain having negative sign was observed for both SGP and NF. However, the nanofiber showed a higher value (44.71) than the SGP (22.63). The calculated $d_{33}$ value for SGP similar to other reports [45]. The piezoelectric strain coefficients are ~1.97 ($d_{33}$) and ~2.66 ($d_{31}$) times higher for the NF than the SGP [Table 4]. It is important to note that such an enhanced $d_{33}$ and $d_{31}$ values for the nanofibers is due to the enhanced flexibility than the SGP.

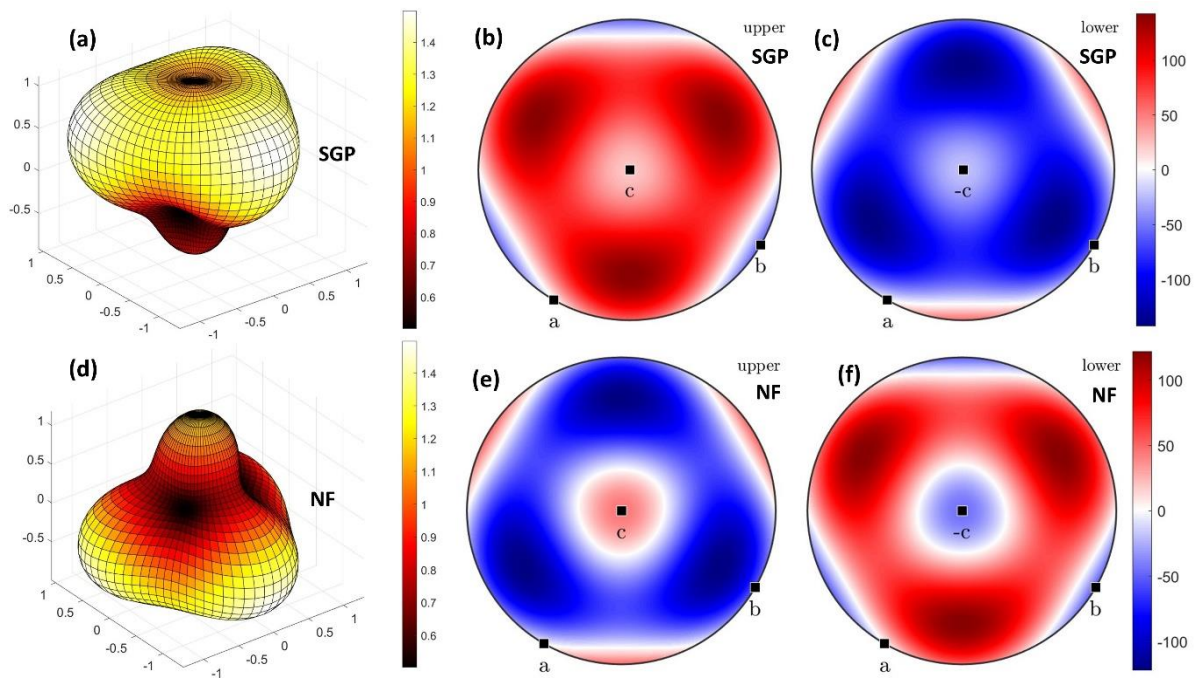

**Figure 9** (a) 3D representation of the piezoelectric strain tensor for SGP (b) 2D representation of the upper half of the sphere for SGP (c) 2D representation of the lower half of the sphere for SGP (d) 3D representation of the piezoelectric strain tensor for NF (b) 2D representation of the upper half of the sphere for NF (c) 2D representation of the lower half of the sphere for NF. [The tensor was plotted in MTEX considering point group 3m (X ∥ a, Y ∥ b, Z ∥ c)].

In a piezoelectric material, applying stress induces uniform polarization across the crystal, generating charges on its surfaces. In a closed circuit, this surface charge movement results in a current flow when stress is applied to the piezoelectric. Hence, as per the direct piezoelectric effect, the electrical polarization vector ($P_i$) is given by: $P_i = d_{ij}\sigma_j$ ($i = 1,2,3; j = 1,2,...,6$), where, $\sigma_j$ is the applied stress [47]. In the case of piezo-phototronic coupling using an ultrasonic cleanser vibration, the catalyst experiences the stress produced by the mechanical vibration. According to Wu et al., ultrasonic wavelengths can produce acoustic pressure with fluctuating waveforms in the range of $10^5$ to $10^6$ Pa [11]. As both the SGP and nanofibers (NF) experienced same stress under ultrasonication, the piezoelectric strain coefficients "$d_{ij}$" is the deciding factor for the NF to generate a larger piezoelectric polarization compared to the SGP. As the $d_{3j}$ elements are much higher for the NFs compared to the SGPs, the piezoelectric polarization along the z-axis will be much higher for the nanofiber. A larger piezoelectric polarization is responsible for a larger inbuilt potential in a piezoelectric material, which improves the charge separation at the surface. Therefore, the higher polarization in the nanofibers facilitate higher charge separation thereby reducing the recombination of the e$^-$-hole pair which ultimately enhances the catalytic performance.

**Conclusion**:

The piezo-phototronic performance was observed to be higher for the electrospun nanofibers as compared with the sol-gel prepared particles under all three illumination conditions of 11W UV, 250W UV, and natural sunlight. Also, the nanofibers showed an enhanced rate constant ($k$) of ~2.5-3.75 times while the sol-gel prepared samples showed only ~1.3-1.4 times higher performance in presence of ultrasonication as compared with only photocatalysis. The improved performance was studied using a first principle study for sol-gel prepared NBT and z-elongated NBT (nanofiber). It was calculated that the elastic moduli ($C_{33}$) for the nanofibers was ~2.15 times lesser for the sol-gel prepared sample. This indicated that the fibers possess higher flexibility than the sol-gel particles. However, the piezoelectric stress coefficients ($e_{31}$ & $e_{33}$) were ~1.2 times lesser for the nanofibers as compared with the sol-gel particle which was the consequence of a lesser Born-effective charge for the fibers. The Born-effective charges revealed that due to the z-elongation in the nanofibers, the covalency in the Bi-O and Ti-O bonds were reduced which reduced the spontaneous polarization in the nanofibers as compared to the sol-gel NBT particles. The consequence was also observed in the reduction of A-O and B-O bond strengths or increase in bond lengths which was studied from XRD. Also, the reduction in the Raman intensity for the Ti-O & TiO$_6$ vibrations were the consequence of decreasing covalency or spontaneous polarization in the nanofibers. However, the piezoelectric strain coefficients increased ~ 2.66 times and ~ 1.97 times for the nanofibers compared with the sol-gel particles. Such an increase was due to the higher flexibility of the fibers that consequently enhanced the piezoelectric polarization in the nanofibers as compared with the sol-gel particles of NBT. This enhanced piezoelectric polarization improved the piezo-phototronic coupling in the nanofibers. These findings suggest that electrospun nanofibers have significant potential for use in applications requiring a combination of mechanical flexibility and efficient piezo-phototronic responses.

**Materials and Methods:**

Synthesis of SGP:

All the precursors used for sol-gel and electrospinning were procured from Alfa-Aesar. For the sol-gel synthesis, sodium nitrate (purity 99.9%) for Na and dihydroxy bis (ammonium lactate) titanium (IV) was dissolved in DI water in separate beakers, while bismuth nitrate pentahydrate (purity 98%) for Bi was dissolved in dilute nitric acid. The individual clear solutions were mixed and stirred until a completely homogeneous solution was formed. The gel was formed by adding citric acid and heating the solution at 80 °C. Further, the gel was burnt using ethylene glycol, which was added to the solution (1:2 ratio of citric acid and ethylene glycol). The burnt powder was denitrified and decarburized at 450 °C for 12 h and 600 °C for 6hr. The phase formation was achieved at 700 °C for 10 h.

Synthesis of NF:

For the electrospinning process, first, a clear solution was prepared. Initially, sodium acetate was dissolved in equal amounts of acetic acid and methoxy ethanol followed by the addition of Bi-nitrate to the solution to achieve a clear solution. An equal amount of acetylacetone was immediately added to the Titanium precursor, Ti-isopropoxide (TTIP, 27.8-28.6%$TiO_2$) in a separate beaker to stabilize the Ti precursor. Then all the individual solutions were mixed in a single beaker to attain a clear solution. The solution was stirred continuously for 12 hours. For the electrospinning process, a viscous solution was prepared by adding 10 ml methanol and 3 g of Polyvinylpyrrolidone (PVP) to 10 ml of prepared solution, followed by continuous stirring at room temperature for 12 hours. Then, the solution was loaded into a syringe for the electrospinning to start. A collector voltage of 25kV was applied by keeping a distance of 10 cm between the needle and drum collector (applied field 2.5kV/cm) and keeping the feed rate at 1ml/hr. The drum collector speed was maintained to be 1000 rpm. A thick white mat-like sample was collected on the aluminium foil, which was further dried at 120 °C for 15 hours to evaporate the solvents. Then, the samples were heated at 400 °C for 1 hour and 700 °C for 1.5 hours for phase formation.

Characterizations:

The X-ray diffraction was performed on the prepared samples using Bruker D2-Phaser (Cu-K$_\alpha$ target) to check the structural information. Full-Prof suite software was used to do the Rietveld refinement of the obtained XRD pattern. The phonon modes were studied from Raman spectroscopy using a Horiba-made LabRAM HR Evolution Raman spectrometer (spectral resolution 1 cm$^{-1}$) having He-Ne LASER of wavelength 633 nm. Fityk was used for the deconvolution of Raman spectra. Diffuse Reflectance Spectroscopy (DRS) study was performed in the range of 200-800 nm (UV-Vis) using a Cary-60, Agilent UV-vis spectrophotometer to obtain the optical band gap of the samples. To check the morphology, Supra55 Zeiss Field Emission Scanning Electron Microscope (FE-SEM) was used. The high-resolution transmission electron microscopy (HR-TEM) was performed using a TFS Talos 200FS system operated at 200 kV. Image J software was used to extract the information from SEM and TEM data. X-ray photoelectron spectroscopy (XPS) data were obtained using a Kratos Axis Supra instrument equipped with a monochromatic Al Kα X-ray source ($h\nu$ = 1486.6 eV) operating at a power of 150 W and under UHV conditions in the range of ~10E$^{-9}$ mbar. All spectra were recorded in hybrid mode, using electrostatic and magnetic lenses and an aperture slot of 300 μm × 700 μm. The wide and high-resolution spectra were acquired at fixed analyser pass energies of 80 $e$V and 20 $e$V, respectively. The samples were mounted in a floating mode to avoid differential charging and thus all spectra were acquired using charge neutralization. The XPS spectra were deconvoluted using XPSPEAK4.1 software. The calibration of the spectra was done using the C-1$s$ peak (Binding energy=284.8 eV). The background extraction was done using the Tougaard function. A combined Gaussian-Lorentzian peak shape was used for all the peaks to get quality fitting. To measure specific surface area and pore volume, Quanta chrome, Autosorb iQ2 BET Surface Area & Pore Volume Analyzer were used. Malvern zeta sizer zs DLS instruments were used to calculate the zeta potential of the prepared samples.

Methylene blue (MB, $C_{16}H_{18}ClN_3S$, SRL chemicals, India) dye was used for the degradation process. A 5ppm MB solution was prepared by adding 5mg MB to 1L DI water and stirring for 1 hour. Then, 50 ml of this stock solution was taken in each beaker and 62.5 mg of SGP and NF were added in respective beakers to prepare a concentration of 1.25 g/L. The prepared solution was stirred for 15 mins. Then the solution was kept undisturbed for 1 hour to ensure proper adsorption-desorption equilibrium of the solute and solvent.

The photocatalysis experiment was done using four low-pressure mercury-vapour (LPMV) discharge UV lamps which emit UV radiation of wavelength ~253.7 nm (Philips TUV 11W PLS) and medium pressure mercury vapour (MPMV) lamp of 250W that emits UV radiation in the range of 200 nm-600 nm. The piezo-photocatalysis was performed using the same light source and an ultrasonic vibration of frequency 40kHz using an ultrasonic cleanser. The intensity of light was ~100lx (0.79 W/m$^2$) which is ~1266 times lesser than the AM 1.5 G sun (1000.4 W/m$^2$). A similar piezo-photo coupling experiment was done for the nanofibers under natural sunlight having a surrounding temperature of ~30-35 °C (~7×10$^4$ lx ~1.56 times lesser than the AM 1.5G sun ~1,09,880 lx). While experimenting, the solution was kept in a water bath to ensure the temperature to not reach beyond 35 °C. The light and sonication were given for 15 mins and then the solution was kept to rest under dark conditions for another 10 mins. Then 5 ml solution was transferred to a centrifuge tube and centrifuged for 15 mins at a rpm of 5000. The centrifuged solution was further used to record the absorbance using the UV-Vis set-up by Researchgate India.

Computational details:

Spin-polarized density functional theory calculations were performed using the Vienna *Ab Initio* Simulation package (VASP). The plane-augmented wave (PAW) method is used for the electron-ion interactions. We have used PBE functionals within the GGA approximations for a better treatment exchange-correlation part. A cutoff of 500 $eV$ is used for the representation of Kohn-Sham wavefunctions, and conjugate-gradient algorithm is used for the structural relaxation and optimization. An energy convergence criterion is set to 10$^{-06}$ $eV$. All the geometric structures are optimized until the Hellmann-Feynman forces on each atom reduces to 0.01 $eV$/angs. A $\Gamma$-centered **k** point grid of $11 \times 11 \times 4$ is used to sample out the Brillouin zone (BZ) for NBT unit cell. To account for the exact electron correlation, the (Heyd-Scuseria-Ernzerhof) HSE06 and metaGGA (SCAN) functionals are employed for the accurate description of electronic properties.

Some *ab initio* simulations were carried out using the VASP code for the calculation of piezoelectric stress coefficient, e$_{ijk}$$^T$ from Density Functional Perturbation Theory (DFPT). For the calculation of mechanical (elastic tensor; $C_{ij}$, compliance tensor; $s_{ij}$ (= $C_{ij}^{-1} \times I$) and piezoelectric properties (piezoelectric tensor; $e_{ijk}^T$, Born effective charges; $Z_{ij}^*$) with ionic contributions, an energy cutoff of 1000 $eV$ was used for accurate description of Hessian matrix

**Acknowledgements**:

The authors KSS and MP would like to acknowledge Ministry of Education, India for providing Prime Minister Research Fellowship (PMRF). The author SS is thankful to the Department of Science and Technology (DST), Govt. of India, for funding through grant no: DST/TDT/AMT/2017/200. The authors also acknowledge the Department of Science and Technology (DST), Govt. of India, New Delhi, India, for providing FIST (SR/FST/PSI-225/2016) instrumentation fund to the discipline of Physics, IIT Indore to purchase a Raman Spectrometer. The authors would like to provide their sincere gratitude towards the Sophisticated Instrument Center (SIC) facilities for providing the FESEM facility at IIT Indore. A. Mekki, K. Harrabi, and S. Sen appreciatively acknowledge the support of the King Fahd University of Petroleum and Minerals, Saudi Arabia, under the DF191055 DSR project.

**Supporting Information:**

The supporting file contains Rietveld refinement fitting and Raman spectra fitting of both SGP and NF samples. The HRTEM images and elemental compositions are also provided. The XPS plots and its detailed explanation is mentioned in the supporting file. The time resolved absorbance spectra for all illumination conditions for both samples are also given. The DFT calculated band diagrams for all compressive and expansion conditions are presented. The as calculated elastic moduli and piezoelectric stress tensors are provided in the supporting file. The calculated Born effective charges are tabulated in the supporting file.

**Conflicts of Interest:**

The authors declare that they have no known competing financial interests or personal relationships that could have appeared to influence the work reported in this paper.

**For Table of Contents Only:**

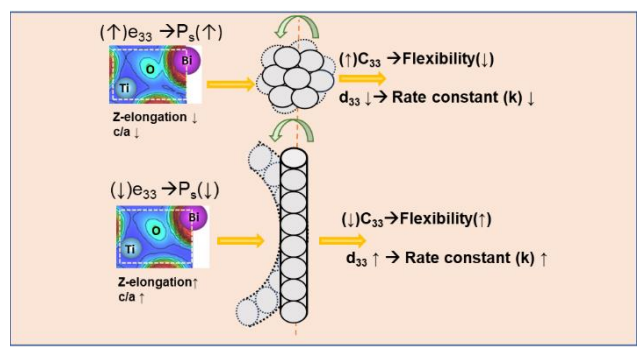